\documentclass[aps,preprint]{revtex4}
\usepackage{amsfonts}
\usepackage{amsmath}
\usepackage{amssymb}
\usepackage{epsf}
\usepackage{epsfig}
\usepackage{graphicx}
\usepackage{rotating}
\usepackage{boxedminipage}

\begin{document}

\title{Critical dynamics of the Potts model: short-time Monte Carlo
simulations. }
\author{Roberto da Silva }
\email{rdasilva@inf.ufrgs.br (corresponding author)}
\affiliation{Departamento de Inform\'{a}tica Te\'{o}rica, Instituto de Inform\'{a}tica,
Universidade Federal do Rio Grande do Sul. \\
Av. Bento Gon\c{c}alves, 9500, CEP 90570-051 Porto Alegre RS Brazil}
\author{J. R. Drugowich de Fel\'{\i}cio }
\email{drugo@usp.br}
\affiliation{Departamento de F\'{\i}sica e Matem\'{a}tica, Faculdade de Filosofia, Ci\^{e}%
ncias e Letras, Universidade de S\~{a}o Paulo.\\
Av. Bandeirantes, 3900-CEP 014040-901 Ribeir\~{a}o Preto SP Brazil}

\begin{abstract}
We calculate the new dinamic exponent $\theta $ of the 4-state Potts model,
using short-time simulations. Our estimates $\theta _{1}=-0.0471(33)$ and $%
\theta _{2}=$ $-0.0429(11)$ obtained by following the behavior of the
magnetization or measuring the evolution of the time correlation function of
the magnetization corroborate the conjecture by Okano et. al. In addition,
these values agree with previous estimate of the same dynamic exponent \ for
the two-dimensional Ising model with three-spin interactions in one
direction, that is known to belong to the same universality class as the
4-state Potts model. The anomalous dimension of initial magnetization $%
x_{0}=z\theta +\beta /\nu $ is calculated by an alternative way that mixes
two different initial conditions. We have also estimated the values of the
static exponents $\beta $ and $\nu $. They are in complete agreement with
the pertinent results of the literature.

\end{abstract}

\maketitle

\setlength{\baselineskip}{0.7cm}

\section{Introduction}

Several results about critical phenomena have been recently obtained using
Monte Carlo simulations in short-time regime \cite{Zheng}. Such simulations
are more convenient than traditional ones done in the equilibrium regime
because they circumvent a known problem in computational physics: the
critical slowing down phenomena.

Investigating Monte Carlo simulations before attaining equilibrium permits
to us obtaining the static critical exponents ($\beta $ and $\nu $) but also
leads to the less known dynamical ones. The scaling equation for
non-equilibrium regime was obtained by Jansen \textit{et al.} \cite{Jansen},
on the basis of renormalization group theory. For systems without conserved
quantities like energy and magnetization (model A in the terminology of
Halperin and Hohenberg \cite{Halperin}), it is written as
\begin{equation}
M^{(k)}(t,\tau ,L,m_{0})=b^{\frac{-k\beta }{\nu }}M^{(k)}(b^{-z}t,b^{\frac{1%
}{\nu }}\tau ,b^{-1}L,b^{x_{0}}m_{0})  \label{Jansen_eq}
\end{equation}%
where $m_{0}$ is the initial magnetization, $x_{0}=$ $x_{0}(m_{0})$ \cite%
{PRL do Zheng} is the anomalous dimension of the initial magnetization, $%
\beta $ and $\nu $ are the known static exponents and $z$ is the dynamic one
($\tau \sim \xi ^{z}$). The values of $M^{(k)}$ are the $k$th moments of
magnetization, defined by $M^{(k)}=\left\langle M^{k}\right\rangle $, where $%
\left\langle \cdot \right\rangle $ indicates an average over several samples
randomly initialized but satisfying the condition of having the same
magnetization $(m_{0})$ at the beggining. The relevance of this result is
allowing to include the dependence on the initial conditions of the dynamic
systems in their non-equilibriun relaxation. As an important consequence,
they could advance the existence of a new critical exponent $\theta $, which
is independent of the known set of static exponents and even of the dynamic
exponent $z$.

By considering large systems and choosing $\tau =0$ $(T=T_{c})$ ; $b^{-z}t=1
$ in the equation (\ref{Jansen_eq}), we obtain the dynamic scaling law for
the magnetization:
\begin{equation}
M(t,m_{0})=t^{\frac{-\beta }{\nu z}}M(1,t^{\frac{x_{0}}{z}}m_{0}).
\label{scaling}
\end{equation}%
Expanding $M$ around the zero value of the parameter $u=t^{\frac{x_{0}}{z}%
}m_{0}$, one obtains:
\begin{equation}
M(t,m_{0})=m_{0}t^{\theta }+O(u^{2})  \label{ordem}
\end{equation}%
which leads to the power law:
\begin{equation}
M(t)\sim m_{0}t^{\theta }  \label{initial slip}
\end{equation}%
whereas $t<t_{0}\sim m_{0}^{-\frac{z}{x_{0}}}$ since in this regime $u\ll 1$%
. This new universal stage, characterized by the exponent $\theta =\left(
x_{0}-\frac{\beta }{\nu }\right) /z$ , has been exhaustively investigated to
confirm theoretical predictions and to enlarge our knowledge on phase
transitions and critical phenomena.

Indeed the anomalous behavior of the magnetization at the beginning of
relaxation, also called critical initial slip, was originally associated to
a positive value of $\theta $. This kind of behavior was confirmed in the
kinetic $q=2$ and $3-$state Potts models \cite{Zheng}, as well as in
irreversible models like the majority voter one \cite{TTome} and the
probabilistic cellular automaton proposed by Tom\'{e} and Drugowich de Fel%
\'{\i}cio to describe part of the immunological system \cite{TD}. However,
the exponent $\theta $ can also be negative. This possibility was observed
in the Blume-Capel model (analytical \cite{oerding} and numerically \cite%
{nossopaperphysrevE}) and in the Baxter-Wu model (numerically \cite%
{Drugoarashiro}), an exactly solvable model which shares with the $4-$state
Potts model the same set of critical exponents. In addition, a negative but
close to zero value for $\theta $\ was obtained for the two-dimensional \
(2-D) Ising model with three-spin interactions in just one direction (IMTSI)
\cite{DrugoSimoes}, a model which is also known to behave to the $4-$state
Potts model universality class.

However, at the best of our knowledge a numerical determination of the
exponent $\theta $ for this special case ($q=4$) of the Potts model
continues to be lacking. Thus, we decided to investigate the short-time
behavior of that model in order to better understand its dynamics and also
to learn about the role of the exponent $x_{0}$ of the nonequilibrium
magnetization concerning the well established classification of the models
(at least in what concerns the static behavior) in the universality classes.

As we mentioned above the estimates for the exponent $\ \theta $ for two
models that belong to the same universality class as the 4-state Potts model
have revealed discrepant results. Whereas the two-dimensional Ising model
with threee spin interactions in one direction (IMTSI) obeys the conjecture
by Okano et al. \cite{Okano}($\theta $ should be negative and close to zero)
\cite{DrugoSimoes} the Baxter-Wu model \cite{Drugoarashiro} strongly
disagrees from that prediction. Putting in numerical values, whereas the
IMTSI model exhibits a small and negative value $(-0.03\pm 0.01)$ for $%
\theta $, when studying the BW model we found $\theta =-0.186\pm 0.002$.

In this paper we study the short-time behavior of the two-dimensional
four-state Potts model and present numerical estimates for $\theta $ using
two different methods. First, we fix the initial value of the magnetization
and follow its time evolution to obtain $\theta (m_{0})$ which in turn can
be extrapolated to lead to $\theta $ when $m_{0}\rightarrow 0$. Second, we
deal with the time correlation of the magnetization defined by \cite{TTome}:

\begin{equation}
C(t)=\frac{1}{N^{2}}\left\langle M(t)M(0)\right\rangle
\end{equation}%
where the average is done over samples whose initial magnetizations are
randomly choosen but obey $\left\langle M(0)\right\rangle =0$. In reference
\cite{TTome} it was shown that this quantity exhibits at $T=T_{c}$ the
power-law behavior:
\begin{equation}
C(t)\sim t^{\theta }\text{.}  \label{corr}
\end{equation}%
This approach has several advantages when compared to the other technique
but its application was initially restricted to the models which exhibit
up-down symmetry. Recently, Tome \cite{TTomeII} has shown that this result
is more general and can be applied whenever the model presents any group of
symmetry operations related to the Markovian dynamics. Equation \ref{corr}
was shown to be valid for instance in the antiferromagnet ordering model,
models with one absorvent state and even in the Baxter-Wu model \cite%
{Drugoarashiro} which exhibits a $Z(2)\otimes Z(2)$ symmetry.

It is well known that the $q\neq 2$ Potts Model does not have up-down
symmetry. However, based on the paper of Tom\'{e} \cite{TTomeII} we can use
this approach to calculate the exponent $\theta $ of the three- and
four-state Potts models.

We checked this possibility working with the three-state Potts model. Our
estimate is $\theta =0.072(1)$, in good agreement with the result obtained
by Zheng \cite{Zheng} ($\theta =0.070(2)$) and with the estimate obtained by
Brunstein and Tome \cite{Brunstein} using a cellular automaton which
exhibits C3v symmetry, the same as the three-state Potts model. A plot of
the correlation function of the magnetization $C(t)$ in this case is
presented in the Fig. 1.

\begin{figure}[th]
\centerline{\psfig{file=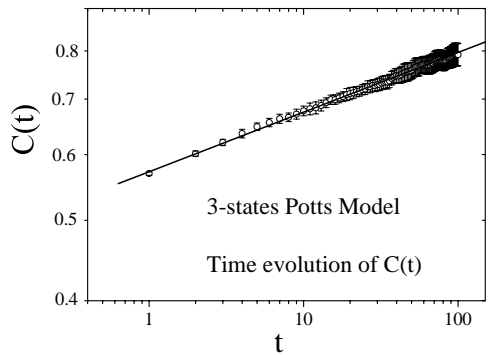,width=8cm}} \vspace*{8pt}
\caption{Log-log plot of $C(t)\times t$ to 3-states Potts Model}
\end{figure}

In the sequence we estimated the exponent $\theta $ of the 4-state Potts
model which agrees with the conjectured by Okano et al. In addition, they
agree with previous estimates for $\theta $ obtained by Sim\~{o}es and
Drugowich de Fel\'{\i}cio \cite{DrugoSimoes}.

To confirm our first estimate we simulated the magnetization evolution for
four different initial values of $m_{0}$ which after extrapolated ($%
m_{0}\rightarrow 0$) led to a similar result.

It is clear that having the exponent $\theta $ the anomalous dimension $%
x_{0} $ follows. But inspired in previous studies \cite{nosso paper Phys.
Letters} we decided to investigate a direct manner in determining $x_{0}$
using mixed initial conditions. In order to build the adequate function we
remember that when in contact with a heat bath at $T=T_{c}$ the
magnetization of completely ordered samples ($m_{0}=1$) decays like a power
law

\begin{equation}
M(t)\sim t^{^{\frac{-\beta }{\nu z}}}\text{.}  \label{magnetization decay}
\end{equation}

Thus, according to the scaling relations (\ref{initial slip}) and (\ref%
{magnetization decay}) it is enough to work with the ratio
\begin{equation}
F_{3}(t,L)=\frac{\left\langle M\right\rangle _{m_{0}\rightarrow 0}}{%
\left\langle M\right\rangle _{m_{0}=1}}  \label{f2}
\end{equation}%
to obtain a power law which decays as $t^{\theta +\beta /\nu z}$ $%
=t^{x_{0}/z}$ . Using $z$ as input \cite{nosso paper Phys. Letters} we can
achieve the value of $x_{0}$. On the other hand, by following the relation (%
\ref{magnetization decay}) we estimate the ratio $\beta /\nu z$ which can be
compared with the exact result $\beta /\nu =1/12$ after using the value of
the dynamical exponent $z$.

Finally using derivatives of the magnetization at an early time and once
more the value of $z$ as input we obtain the critical exponent $\nu $ of the
correlation length, whose numerical estimates by usual techniques
(phenomenological renormalization group, hamiltonian studies and equlibrium
Monte Carlo simulations) are always very different from the pertinent result
($\nu =2/3$).

The paper is organized as follows: in the next section we present a brief
review and some details of the simulation. The results are presented in
Section 3 and our conclusions are in section 4.

\section{ The kinetic $4$-states Potts model}

In time-dependent simulations we are interested in finding power laws for
physical quantities even when the system is far from equilibrium. In this
regimen the magnetization $M(t)$ must be calculated as an average over
several samples because the system does not obey any \textit{a priori }%
probability distribution. The average can be done in different ways: we can
prepare all the samples with the same initial magnetization (sharp
preparation) or generate samples which satisfy a less restrict criteria like
to have mean value of the magnetization equal to zero.

The $q$ states Potts model ferromagnetic without the presence of an external
field is defined by the Hamiltonian \cite{potts}:

\begin{equation}
H=-J\sum \limits_{\left\langle i,j\right\rangle
}\delta_{\sigma_{i},\sigma_{j}}  \label{hamiltoniana}
\end{equation}
where $J$ denotes the interaction between the nearest neighbors $%
\left\langle i,j\right\rangle $ and $\sigma_{i}$ can assume different values
$\sigma _{i}=0,1..,q-1$. If two spins are parallel they contribute with
energy $-J$, else the energy is null.

The critical temperature of this model, is known exactly \cite{potts} ,
\begin{equation}
\frac{J}{k_{B}T}=\log[1+\sqrt{q}].  \label{parcritico}
\end{equation}

The magnetization, different of other models as Ising and Blume Capel is not
only the sum of variables of spin. A general expression used for the
magnetization in the Potts model that considers an average over the sites
and over the samples is written as
\begin{equation}
\left\langle M(t)\right\rangle =\frac{1}{N_{s}(q-1)N}\sum%
\limits_{j=1}^{N_{s}}\sum\limits_{i=1}^{L^{d}}(q\delta _{\sigma
_{i,j}(t),1}-1)  \label{mag}
\end{equation}%
where $\sigma _{i,j}$ denotes the spin $i$ of the $j$th. sample at the $t$%
th. MC sweep. Here $N_{s}$ denotes the number of the samples and $L^{d}$ is
the volume of the system. This kind of simulation is performed $N_{B}$ times
to obtain our final estimates as a function of $t$. In order to prepare a
lattice with a given magnetization we need to choose the states $\sigma
_{i}=0,1,2,3$ in each site with equal probability (1/4). Next we measure the
magnetization and change states in sites randomly chosen in order to obtain
a null value for the magnetization. Finally we change $\delta $ sites
occupied by $\sigma _{i}=0,2$ or $3$ and substitute by $\sigma _{i}=1$. The
initial magnetization in this case will be given by:
\begin{equation}
m_{0}=\frac{4\delta }{3N}  \label{initial_magnetization}
\end{equation}%
where $N=L^{d}$.

We have chosen to update the spins according to the heat bath algorithm,
which means that the probability that the spin $\sigma _{\overrightarrow{x}}$
localized at the site $\overrightarrow{x}$ can assume another value $\sigma
_{\overrightarrow{x}}^{f}$ is given by:
\begin{equation}
P(\sigma _{\overrightarrow{x}}\rightarrow \sigma _{\overrightarrow{x}}^{f})=%
\frac{\exp(-\frac{J}{k_{B}T}\sum\limits_{\overrightarrow{k}}\delta _{\sigma
_{\overrightarrow{x}+\overrightarrow{k}}(t),\sigma _{\overrightarrow{x}%
}^{f}})}{\sum\limits_{\sigma _{\overrightarrow{x}}^{f}=0}^{3}\exp(-\frac{J}{%
k_{B}T}\sum\limits_{\overrightarrow{k}}\delta _{\sigma _{\overrightarrow{x}+%
\overrightarrow{k}}(t),\sigma _{\overrightarrow{x}}^{f}})}  \label{prob}
\end{equation}%
where $\sum\limits_{\overrightarrow{k}}$ denotes the sum about the nearest
neighbors sites of the spin at the site $\overrightarrow{x}$ .

\section{Results}

We performed Monte Carlo simulations for $N_{B}=5$ different bins and four
different initial magnetizations, $m_{0}=4\delta /3N\approx 0.033$, $0.049$,
$0.066$ and $0.082$, for a lattice of size $L=90$. These magnetizations
correspond respectively to $\delta =200$, $\delta =300$, $\delta =400$, $%
\delta =500$ according to the expression (\ref{initial_magnetization}). To
estimate the errors we ran $5$ different bins, with $35000$ samples each one.

\begin{figure}[th]
\centerline{\psfig{file=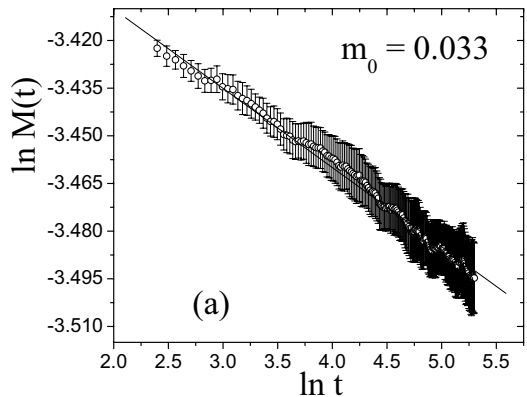,width=8cm}} \vspace*{8pt}
\caption{Time evolution of the magnetization for $m_{0}=0.033$.}
\end{figure}

In the figs. 2 and 3 we present the log-log plot for two different initial
magnetizations and two different values of $m_{0}=0.033$ and $m_{0}=0.082$
respectively. The fit had excellent goodness $(Q=0.99)$ in the entire range $%
[t_{i},t_{f}]\subset \lbrack 0,200]$. We have chosen the range $[10,100]$ to
estimate the value to $\theta $ for all initial magnetizations. We must note
that the slope is positive for $m_{0}=$ $0.082$\ \ but changes the signal at
some value between \ $0.066$ \ and $0.049$. This trend to a negative value
can be also certified in the Fig. 4 that shows the plot of $\theta =\theta
(m_{0})$. It is worth to mention that there is a clear linearity in the plot
of $\theta $ versus the initial magnetization (see fig. 4) but big
fluctuations are observed in the evolution of the magnetization in each
case.The value found for $\theta _{ext}$ is $-0.0471(33)$ is in fair
agreement with the result \cite{DrugoSimoes} for the IMTSI model and
strongly disagrees with the result obtained for the Baxter-Wu model.

\begin{figure}[th]
\centerline{\psfig{file=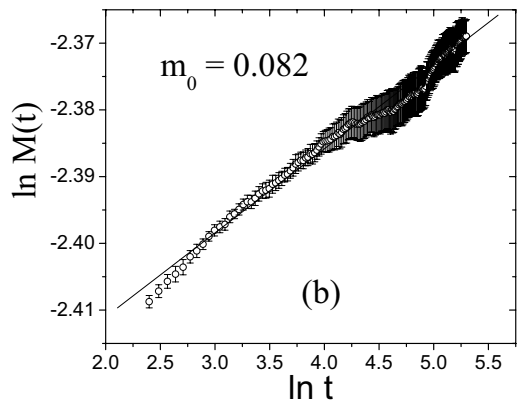,width=8cm}} \vspace*{8pt}
\caption{Time evolution of the magnetization for $m_{0}=0.082$.}
\end{figure}

We also ran Monte Carlo simulations to obtain the time-evolution correlation
$C(t)$, and the the dynamic exponent $\theta $ according to the relation (%
\ref{corr}). For the same interval $[10,100]$ we have gotten the estimate $%
\theta =$ $-0.0429(11)$ that corroborates the above estimate $-0.0471(33)$
obtained from the evolution of the non-equilibrium magnetization. We used $%
35000$ samples and once more 5 different bins to make this estimate. In
figure 5 it is shown the log-log plot of $C(t)\times t$ for the four-state
Potts model.

\begin{figure}[th]
\centerline{\psfig{file=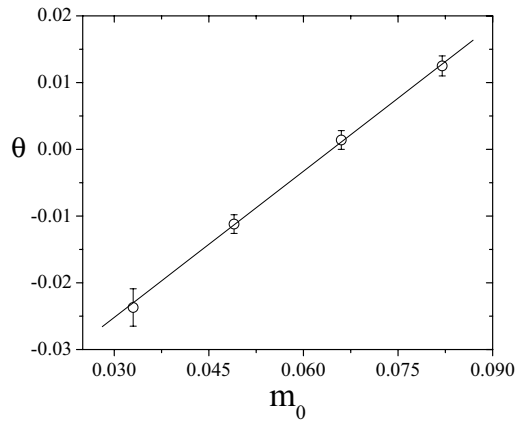,width=8cm}} \vspace*{8pt}
\caption{$\protect\theta $ as function of $m_{0}$ in the 4-state Potts Model
(extrapolation)}
\end{figure}

Equilibrium simulations for the four-state Potts model are ever acompannied
by strong fluctuations. The reason is the presence in the Hamiltonian of a
marginal operator characterized by an anomalous dimension equal to zero. If
we hope some signal of that anomalous behavior in short-time simulations
done far from equilibrium we should look for that in the only new anomalous
dimension: that of the initial magnetization $x_{0}$. In order to confirm
the null value for the anomalous dimension we estimate directly $x_{0}$
using the reported function $F_{3}$.
\begin{figure}[th]
\centerline{\psfig{file=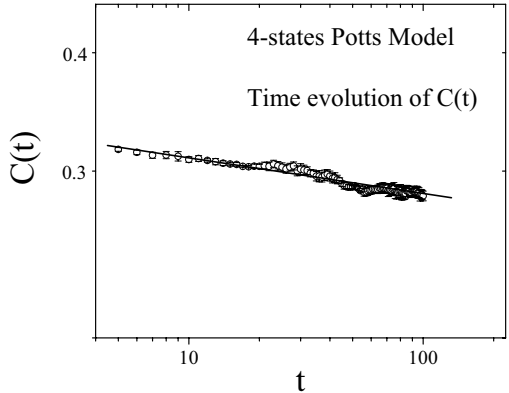,width=8cm}} \vspace*{8pt}
\caption{Log-log plot of $C(t)\times t$ to 4-states Potts Model}
\end{figure}
Over here we used $N_{b}=5$ bins for each initial condition and again the $%
N_{s}=35000$ realizations. A plot when the initial magnetization is $%
m_{0}=0.082$ can be seen in the fig. 6. Extrapolating $\ $\ to $%
m_{0}\rightarrow 0$ we found $\left( x_{0}/z\right) _{ext}=0.0077(33)$ with
the same set initial magnetizations $m_{0}$ that was used to extrapolating $%
m_{0}$. Using as input $z=2.290(3)$, we obtain $x_{0}=0.0176(75)$.

The value of $\beta /\nu $ can be obtained directly of magnetization decay
of a initial ordered state, considering the relation:
\begin{equation}
\frac{\beta }{\nu }=\left( \frac{\beta }{\nu z}\right) _{c}\cdot z\text{.}
\label{betani}
\end{equation}%
The value of the exponent obtained from the decay $(M\sim t^{-\frac{\beta }{%
\nu z}})$, was $\left( \frac{\beta }{\nu z}\right) _{c}=0.0547(1)$ with $%
Q=0.84$.
\begin{figure}[th]
\centerline{\psfig{file=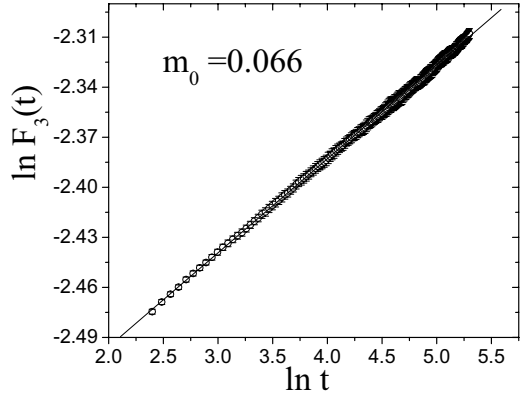,width=8cm}} \vspace*{8pt}
\caption{Plot of $F_{3}$ function for case $m_{0}=0.066.$}
\end{figure}
Using as the input the same value for $z$ value we achieved $\frac{\beta }{%
\nu }=0.12526(28)$ that is in a good agreement with the exact result $\frac{%
\beta }{\nu }=0.125$ \cite{potts}.

Finally to find the exponent $\nu $, we calculated the expected decay of
derivative

\begin{equation}
\begin{tabular}{lll}
$D(t)$ & $=$ & $\left. \dfrac{\partial}{\partial\tau}\ln
M(t,\tau)\right\vert _{\tau=0}$ \\
\  & \  & \  \\
\  & $\underset{=}{N}$ & $\dfrac{\ln M(t,\tau+\delta)-\ln M(t,\tau-\delta )}{%
2\delta}.$%
\end{tabular}
\   \label{derivative}
\end{equation}

According to the reference (\ref{Jansen_eq}), for $m_{0}=1$, we expect of $%
D(t)$ the power law behavior%
\begin{equation}
D(t)\sim t^{\frac{1}{\nu z}}  \label{lei de potencia derivada}
\end{equation}%
and so $\nu =$ $\left( \dfrac{1}{\nu z}\right) _{c}^{-1}\dfrac{1}{z}$. In
the Fig. 7 we illustrate the power law (\ref{lei de potencia derivada}),
performing simulations with $N_{s}=8000$ samples with $N_{b}=5$ to estimate
errors and $t_{\max }=1000$ MC steps.
\begin{figure}[th]
\centerline{\psfig{file=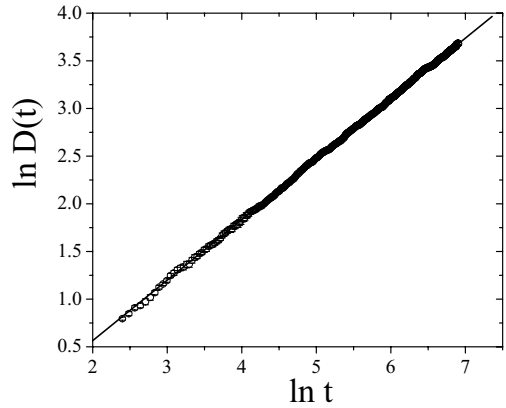,width=8cm}} \vspace*{8pt}
\caption{Time dependence of derivative $D(t)$ of the 4-state Potts Model.
The slope gives $1/\protect\nu z$.}
\end{figure}
Our best estimative is $\dfrac{1}{\nu z}=0.65453(46)$, in the interval $%
[10,1000]$ with $Q=0.99$. The value found to $\nu $ at this interval is $\nu
=0.667(1)$. The value $\left( \dfrac{1}{\nu z}\right) _{c}$ also is used to
estimate $\beta $. This minimizes the errors because:%
\begin{equation}
\beta =\left( \frac{\beta }{\nu z}\right) _{c}\left( \dfrac{1}{\nu z}\right)
_{c}^{-1}  \label{beta}
\end{equation}

The value found to $\beta $ is $0.0836(2).$ These \ values are in complete
agreement with conjectured results $\nu =2/3=0.\overline{6}$ and $\beta
=1/12=0.08\overline{3}$.

\section{Conclusions}

We calculated the dynamic exponent $\theta $ of the four-state Potts Model
using two different techniques: the time-evolution of the magnetization when
the samples are prepared with a small and nonzero value of $m_{0}$ for four
different values of initial magnetization and through of the time
correlation of the magnetization. The values found for $\theta $ in both
cases corroborate the conjecture by Okano et al. and are in complete
agreement with the estimate obtained for another model which is in the same
universality class as the 4-state Potts model: the 2-D Ising model with
three-spin interactions in one direction. We also estimated directly the
anomalous dimension of the initial magnetization $x_{0}$ which results very
closed to zero. The same does not occur in the Baxter-Wu model that,
although exhibiting the same leading exponents as the 4-state Potts model,
does not own a marginal operator which prevents us from determine with good
precision the critical exponents.

We also estimated the static exponents $\beta $ and $\nu $ using the
short-time Monte Carlo simulations and our results are in surprisingly
agreement with pertinent results.

As an additional contribution we present a new estimate for the exponent $%
\theta $ of the 3-state Potts Model working with the time evolution of the
time correlation of the magnetization $C(t)$. This kind of approach has been
previoulsy used by Brunstein and Tom\'{e} when studying a four-state
cellular automaton with C-3$\nu $ symmetry \cite{Brunstein}.The result
agrees with estimates obtained from the evolution of the magnetization \cite%
{Zheng} after the extrapolation ($m_{0}\rightarrow 0$).

\section*{\textbf{Acknowledgments}}

The authors thank to CNPq for financial support.

\end{document}